\documentclass[12pt]{article}
\usepackage{graphicx}
\usepackage[bookmarksnumbered,colorlinks,plainpages]{hyperref}

\begin{document}

\title{How to proceed with nonextensive systems at thermally stationary state?}

\author{Q.A. Wang$^a$, L. Nivanen$^a$, M. Pezeril$^b$, and A. Le M\'ehaut\'e$^a$ \\
$^a$Institut Sup\'erieur des Mat\'eriaux du Mans, \\
44, Avenue F.A. Bartholdi, 72000 Le Mans, France \\
$^b$Laboratoire de Physique de l'\'etat Condens\'e, \\ Universit\'e du Maine, 72000 Le Mans,
France}

\date{}

\maketitle

\begin{abstract}
In this paper, we show that 1) additive energy is not appropriate for discussing the validity
of Tsallis or R\'enyi statistics for nonextensive systems at meta-equilibrium; 2) $N$-body
systems with nonadditive energy or entropy should be described by generalized statistics whose
nature is prescribed by the existence of thermodynamic stationarity. 3) the equivalence of
Tsallis and R\'enyi entropies is in general not true.
\end{abstract}

{\small PACS : 02.50.-r, 05.20.-y, 05.70.-a}

\vspace{1cm}

\section{Introduction}
Although scientists apply Boltzmann-Gibbs statistics (BGS), or its logarithmic microcanonical
entropy $S=\ln W$ ($W$ is the phase space volume, Boltzmann constant $k=1$) and exponential
probability distributions $p\propto exp(-\beta H)$ ($H$ is Hamiltonian), to systems having
long range interaction or finite size\cite{Conform,Cardy,Janco,Ruff95,Ruff01,Ruff02,Gross},
this classical statistical theory, from the usual point of view, remains an additive theory in
the thermodynamic limits, i.e., the extensive thermodynamic quantities are proportional to its
volume or to the number of its elements. However, the systems having finite size or containing
long range interaction may have nonextensive and nonadditive energy or entropy\footnote{A
clear discussion of these two concepts is given in \cite{Touchette}}. Hence the applications
of BGS to these systems have led to a belief that $S=\ln W$ or $exp(-\beta H)$ is universal,
at least for systems at thermodynamic equilibrium\cite{Gross1}.

During the last years, the development of a nonextensive statistical mechanics (NSM) proposed
by Tsallis\cite{Tsal88} intensified this debate. Polemics take place within NSM to decide
whether or not one should use nonadditive energy with nonadditive entropy and whether a
nonextensive theory should be based on the independence of subsystems of $N$-body
systems\cite{Abe,Toral,Raggio,Beck01,Wang02,Wang02a,Wang02b,Wang02c,Wang02d}. The reader will
find that these problematics are tightly related to the self-consistence and the validity of
the theory for systems at stationary state\footnote{We would like to indicate here that,
according to the actual understanding\cite{Letters}, NSM applies only to non-equilibrium
systems, but from the theoretical point of view, the exact discussions of the formal structure
of NSM, of the zeroth law of thermodynamics, of heat and work and of meta-equilibrium state
within NSM are formally consistent with the principles of the equilibrium thermodynamics.}. On
the other hand, it is just partially due to some fundamental problems of the theory during its
development that the new nonextensive statistics has met reticence among many
physicists\cite{Gross1,Cho,Nauenberg}.

Very recently, arguments\cite{Gross1,Nauenberg} have been forwarded to say that Tsallis
entropy should be rebuffed for equilibrium nonextensive systems. This affirmation is based on,
among others, the works\cite{Abe,Toral} using additive energy to define equilibrium (or meta
equilibrium) and temperature which leads to the equivalence of the nonextensive Tsallis
entropy and the extensive R\'enyi one. In this paper, we would like to show that : 1) additive
energy should be considered as an approximation and is not appropriate for discussing
fundamental questions such as the validity of NSM; 2) stationary $N$-body systems with
nonadditive energy or entropy should be, in principle, described by generalized statistics
whose composition nature is prescribed by the existence of thermodynamic stationarity. The
nonextensive statistics may be Tsallis or R\'enyi one which is naturally associated with
nonadditive energy.

\section{Some consequences of additive energy}
Additive energy formalism of NSM appeared with the study of thermodynamic stationarity and of
zeroth law\cite{Abe,Toral,Raggio} within the third version of NSM using escort
probability\cite{Penni}. This formalism is actually more and more accepted in NSM, even for
the fundamental derivation of Tsallis statistics from first principles for finite
systems\cite{Almeida}.

One of the important arguments for rejecting Tsallis entropy\cite{Tsal88}
\begin{equation}                                    \label{1}
S^T=\frac{W^{1-q}-1}{1-q}
\end{equation}
from the study of microcanonical systems {\it at equilibrium} is based on the results obtained
by using additive energy $E(A+B)=E(A)+E(B)$ for two noninteracting subsystems $A$ and $B$ of a
composite system $A+B$ satisfying joint probability $p(A+B)=p(A)p(B)$ (or $W(A+B)=W(A)W(B)$
for microcanonical ensemble) (as in BGS, this additivity seems justified by the product joint
probability which implies independence of $A$ and $B$). Only under this condition, one gets an
explicit entropy nonadditivity
\begin{equation}                                    \label{2}
S^T(A+B)=S^T(A)+S^T(B)+(1-q)S^T(A)S^T(B).
\end{equation}
Then, if $A+B$ is isolated, thermal equilibrium can be reached with
$\beta(A)=\beta(B)$\cite{Abe}. Here the inverse temperature $\beta$ is given by
\begin{equation}                                    \label{2a}
\beta=\frac{1}{1+(1-q)S^T}\frac{\partial S^T}{\partial E}.
\end{equation}
Toral et al\cite{Toral,Velazquez} indicated that this temperature was identical to that
within Boltzmann thermo-statistics, i.e.
\begin{equation}                                    \label{3}
\beta=\frac{1}{1-q}\frac{\partial \ln [1+(1-q)S^T]}{\partial E}=\frac{\partial \ln
W}{\partial E}=\frac{\partial S}{\partial E}
\end{equation}
for microcanonical ensemble. So in this case, the physically significant entropy is the
Boltzmann one instead of Tsallis one. Eqs.(\ref{1}) and (\ref{2}) turn out to be useless, as
noticed by Gross\cite{Gross1}.

Toral's result can be extended to canonical ensemble\cite{Abe3} with
\begin{equation}                                    \label{4}
S^T=-\frac{1-\sum_ip_i^q}{1-q}
\end{equation}
satisfying Eq.(\ref{2})\cite{Tsal88}, where $p_i$ is the probability that the system is
at the state labelled by $i$. We can see :
\begin{equation}                                    \label{5}
\beta=\frac{1}{1-q}\frac{\partial \ln [1+(1-q)S_q]}{\partial E}=\frac{1}{1-q}\frac{\partial
\ln \sum_ip_i^q}{\partial E}=\frac{\partial S^R}{\partial E}
\end{equation}
where
\begin{equation}                                    \label{6}
S^R=\frac{\ln \sum_ip_i^q}{1-q}
\end{equation}
is R\'enyi entropy\cite{Reny66} which is additive $S^R(A+B)=S^R(A)+S^R(B)$ if the product
joint probability holds. So it seems that, for equilibrium or stationary canonical systems,
Tsallis nonadditive entropy is equivalent to R\'enyi additive one\cite{Tsal02}. In addition,
for microcanonical ensemble, if we suppose complete probability distribution with
$\sum_{i=1}^wp_i=1$, we have
\begin{equation}                                    \label{7}
S^R=\ln w
\end{equation}
since $\sum_{i=1}^wp_i^q=w^{1-q}$ where $w$ is the total number of states of the system.
So R\'enyi entropy and Boltzmann one are equivalent for microcanonical ensemble. This is
consistent with the reduction of Tsallis entropy to Boltzmann one for microcanonical
ensemble.

There are other examples\cite{Abe2,Vives} of this self-reduction of Tsallis nonadditive
statistics to additive statistics due to additive energy, e.g. the study of ideal
gas\cite{Abe2} within the third version of NSM, the internal energy $U_q$ is defined by
$U_q=\frac{\sum_ip_i^qE_i}{\sum_ip_i^q}$ , where
\begin{equation}                                    \label{8a}
p_i\propto [1-(1-q)\beta(E_i-U_q)]^{1/(1-q)}
\end{equation}
is the generalized canonical distribution and $E_i$ is the energy of the state $i$. This
generalized internal energy of a, say, nonextensive ideal gas turns out to be
additive\cite{Abe2} :
\begin{equation}                                    \label{8}
U_q=\frac{3N}{2\beta}
\end{equation}
which is identical to the Boltzmann ideal gas and completely independent of the
nonadditivity $q$.

Now the question is whether or not this self-reduction of NSM to other statistics of additive
entropy due to additive energy arises systematically in all applications? In one of our recent
papers presenting a general analysis of the third version of NSM, it was shown\cite{Wang02}
that, through a series theoretical anomalies, Tsallis statistics might be mathematically
self-consistent and physically operational only when $q=1$ if one use the temperature defined
in Eq.(\ref{2a}) with additive energy.

\section{Tsallis entropy with additive energy}

Due to the fact that $S^R$ is a monotonically increasing function of $S^T$\footnote{This can
be illustrated by the following relationship : $dS^R=\frac{dS^T}{1+(1-q)S^T}
=\frac{dS^T}{\sum_ip_i^q}$ where $\sum_ip_i^q$ is always positive. This fact should be taken
into account in the study of thermodynamic stability.}, they will reach the extremum together.
One can hope that the maximum entropy (for $q>0$) will give same results with same
constraints. Indeed, R\'enyi entropy has been used to derive, by maximum entropy method, the
Tsallis $q$-exponential distribution Eq.(\ref{8a}) with additive energy and the temperature
given in Eq.(\ref{2a})\cite{Lenzi,Bashkirov}.

As a matter of fact, this equivalence of $S_q$ with $S^R$ or $S$ for equilibrium or stationary
systems is true only for additive energy. It is not true in general for nonextensive systems
with nonadditive energy. And more, this equivalence reveals in fact that the invalidity of the
additive energy formalism of NSM, since it has been established on the basis of the
nonadditivity of $S^T$ in Eq.(\ref{2}) and the additivity of $S^R$ given by
$S^R(A+B)=S^R(A)+S^R(B)$. However, with additive energy, these relationships are no more
valid. Let us see this first for Tsallis entropy.

$S^T$ is associated with the $q$-exponential distributions. For complete
distribution\cite{Tsal88} (the following calculation is also valid for other formalisms of
NSM), the probability of $A+B$ for a joint state $ij$ is :
\begin{eqnarray}                                  \label{9}
p_{ij}(A+B) &=& \frac{1}{Z(A+B)}[1-(q-1)\beta(E_i(A)+E_j(B))]^{1/(q-1)} \\ \nonumber &=&
p_i(A)p_{j\mid i}(B\mid A)
\end{eqnarray}
where $p_i(A)=\frac{1}{Z(A)}[1-(q-1)\beta E_i(A)]^{1/(q-1)}$ is the probability for $A$ to be
at the state $i$ and $p_{j\mid i}(B\mid A)=\frac{1}{Z_i(B\mid A)}[1-(q-1)\beta e_{j\mid
i}(B\mid A)]^{1/(q-1)}$ is a kind of conditional probability for $B$ to be at a state $j$ with
energy $e_{j\mid i}(B\mid A)=E_j(B)/[1-(q-1)\beta E_i(A)]$ if $A$ is at $i$ with energy
$E_i(A)$. In this case, the total entropy is given by
\begin{eqnarray}                                  \label{10}
S^T(A+B) &=& \frac{\sum_ip_i(A)^q\sum_jp_{j\mid i}(B\mid A)^q-1}{1-q} \\ \nonumber &=&
S^T(A)+\sum_ip_i(A)^qS_i^T(B\mid A)
\end{eqnarray}
where $S_i^T(B\mid A)=\frac{\sum_jp_{j\mid i}(B\mid A)^q-1}{1-q}$. This relationship is
totally different from Eq.(\ref{2}). With Eq.(\ref{10}), the discussion of thermodynamic
equilibrium and of zeroth law and the definition of the temperature in Eq.(\ref{2a}) are
impossible. So there is no equivalence between $S^T$ and $S^R$ here.

As a matter of fact, comparing Eq.(\ref{9}) to the product joint probability which still
holds, one gets $p_{j\mid i}(B\mid A)=p_j(B)$ or $e_{j\mid i}(B\mid A)=E_j(B)$ which holds
only when $q=1$!

In the same way, it can be shown that R\'enyi entropy is not additive. So that the definition
of the temperature $\beta=\frac{\partial S^R}{\partial E}$ does not exist. In fact, using the
$q$-exponential distribution associated with $S^R$\cite{Wang03a}, it can be shown that, within
the complete probability formalism, $S^R$ is additive if and only if :
\begin{eqnarray}                                  \label{11}
E(A+B)=E(A)+E(B)+(q-1)\beta E(A)E(B)
\end{eqnarray}
which is also necessary for the nonadditivity of $S^T$ given by Eq.(\ref{2}).

So we see that the condition of additive energy of noninteracting systems is not
appropriate for nonextensive systems implying {\it interacting subsystems} and described
by Tsallis or R\'enyi entropies. If the subsystems are independent, one should simply
return to additive statistics.

However, a paradox seems to arises. From the usual point of view, dependent subsystems and
nonadditive energy do not allow the product joint probability. Without this joint probability,
the $N$-body distribution cannot be related to one-body distribution and the explicit
nonadditivity Eq.(\ref{2}) of Tsallis entropy will disappear. There would be no temperature
and thermodynamic relations. In what follows, we would like to propose a plausible way to
establish nonextensive statistics on the basis of Tsallis entropy for {\it systems at
thermodynamic stationary state} having nonadditive energy, without imposing first of all the
product joint probability which turns out to be intrinsic to the formalism and independent of
whether or not energy is additive.

\section{How to proceed with nonextensive systems at equilibrium}
It is well known that the total hamiltonian is not the sum of the Hamiltonians of
subsystems if there is interaction between $A$ and $B$ or if the system has finite size.
We should write
\begin{equation}                            \label{12}
H(A)=H(A)+H(B)+f_H(A,B).
\end{equation}
In general, if the nonadditive term $f_H(A,B)$ is not known, no exact physical treatment
will be possible. But there exist many effective approach for solving the problem with
empirical parameters. In fact, the approach of NSM concerning $f_H(A,B)$ is a little
special.

\subsection{Tsallis statistics}

In our opinion, the starting point of NSM is to suppose $$f_Q(A,B)=f_{\lambda_Q}[Q(A),Q(B)],$$
where $Q$ is certain physical quantity, $f_{\lambda_Q}$ is a function depending on a constant
$\lambda_Q$ for $Q$. In other words, the nonextensive term of a quantity is uniquely
determined by the same quantity of each subsystem. This choice may have its limits. But the
advantage is to allow a more general formalism of statistics which may formally parallel BGS
and enjoy its mathematical methods.

An interesting method\cite{Abe5} to determine $f_{\lambda_Q}$ is to consider the thermodynamic
equilibrium or stationarity as a constraint on the form of $f_{\lambda_Q}$, i.e., one looks at
systems at equilibrium or stationary states. It is shown\cite{Wang02a,Abe5} that for the
equilibrium or stationarity to take place, we can have
\begin{eqnarray}                            \label{13}
Q(A) &=& \frac{h(A)-1}{\lambda_Q} \\ \nonumber Q(B) &=& \frac{h(B)-1}{\lambda_Q} \\
\nonumber Q(A+B) &=& \frac{h(A+B)-1}{\lambda_Q}
\end{eqnarray}
and $h(A+B)=h(A)h(B)$, where $h(A)$ or $h(B)$ is the factor depending on $A$ or $B$ in the
derivative $\frac{\partial Q(A+B)}{\partial Q(B)}$ or $\frac{\partial Q(A+B)}{\partial Q(A)}$.
For entropy S\cite{Abe5}, this leads to
\begin{equation}                            \label{14}
S(A+B)=S(A)+S(B)+\lambda_SS(A)S(B),
\end{equation}
and for energy\cite{Wang02a}, we get
\begin{equation}                            \label{15}
E(A+B)=E(A)+E(B)+\lambda_EE(A)E(B).
\end{equation}

Now let us see the microcanonical ensemble. From Eqs.(\ref{13}) related to entropy, if we want
that the nonadditive entropy is an extension of Boltzmann one, i.e., it recovers $\ln W$
whenever $\lambda_S=0$, then the simplest choice is $h=W^{\lambda_S}$ giving Eq.(\ref{1}),
i.e., Tsallis entropy with $\lambda_S=1-q$. This leads to $W(A+B)=W(A)W(B)$, the product joint
probability, without supposing the independence of subsystems $A$ and $B$.

For canonical ensemble, we require that $h$ be a trace form function of $p_i$ and that
$S$ recover Boltzmann-Gibbs-Shannon entropy $S=\sum_ip_i\ln (1/p_i)$ for $\lambda_S=0$,
then a simple choice is $S^T=\sum_ip_i\frac{(1/p_i)^{\lambda_S}-1}{\lambda_S}$. This is
Tsallis entropy with $\lambda_S=1-q$. This leads to $p_{ij}(A+B)=p_i(A)p_j(B)$ without
supposing noninteracting system and additive energy, as discussed in
\cite{Wang02b,Wang02c,Wang02d}.

Then the following formal systems of NSM are well known. We can establish NSM in a
coherent way with either the complete distribution $\sum_ip_i=1$\cite{Tsal88} or the
incomplete distribution $\sum_ip_i^q=1$\cite{Wang02,Wang01} with well defined temperature
and forces\cite{Wang02c} according to the nonadditive energy Eq.(\ref{15}). Here we
indicate only that the temperature should be defined by
\begin{equation}                                \label{16}
\beta=Z^a\frac{\partial S^T}{\partial E}
\end{equation}
where $Z$ is the partition function associated with the $q$-exponential distribution
$exp_q(-\beta E)=[1-a\beta E]^{1/a}$ where $a=1-q$ with $\sum_ip_i^q=1$ and $a=q-1$ with
$\sum_ip_i=1$.

\subsection{R\'enyi statistics}
Above approach also applies for R\'enyi statistics for interacting systems with
nonadditive energy\cite{Wang03a}. We consider a more general pseudo-additivity required
by thermodynamic equilibrium\cite{Abe5}
\begin{equation}                                \label{16a}
H[Q(A+B)]=H[Q(A)]+H[Q(B)]+\lambda_Q H[Q(A)]H[Q(B)],
\end{equation}
where $H$ is certain differentiable function satisfying $H(0)=0$. For R\'enyi statistics, let
us put $H[S]=\frac{e^{(1-q)S}-1}{1-q}$ which assures the additivity of R\'enyi entropy
$S^R(A+B)=S^R(A)+S^R(B)$ for $\lambda_S=1-q$. This means that $S^R$ satisfies the requirement
of the existence of equilibrium. The concomitant statistics with nonadditive energy in
Eq.(\ref{15}) is discussed in detail in \cite{Wang03a}. The temperature within this
nonextensive statistics is given by
\begin{equation}                                    \label{17}
\beta= [1+(1-q)\beta E]\frac{\partial S^R}{\partial E}
\end{equation}
or
\begin{equation}                                    \label{18}
\frac{1}{\beta}=\frac{\partial E}{\partial S^R}-(1-q)E.
\end{equation}
Since $[1+(1-q)\beta E]$ is always positive ($q$-exponential probability cutoff), $\beta$ has
always the same sign as $\frac{\partial S^R}{\partial E}$. We see that this temperature has
nothing to do with the one defined with $S^T$. The equivalence between these two entropies via
thermodynamic equilibrium based on additive energy is not exact.

We would like to mention here that R\'enyi entropy has been shown\cite{Lesche,Abe02x} to be
non-observable because an arbitrarily small variation $\delta$ in probability distribution may
lead to an important variation in $S^R$. It should be clear that this conclusion is reached
under the condition\cite{Lesche} that the total number of states $w$ is infinite and $(1/w)^q$
is small compared to $\delta^q$, the small variation of probability. This is a very harsh
condition if we consider that $\delta$ must be arbitrarily small for observability
condition\cite{Lesche}. It should be noted that the asymptotic behavior of $\Delta
S^R(\delta,w)/S_{max}$ for finite $\delta$ and $w\rightarrow \infty$ is different from the one
for arbitrarily small $\delta$ and arbitrary $w$. This second asymptotic behavior should be
more general to our opinion because it applies for any system. Taking the probability
distributions proposed by Lesche\cite{Lesche} and making the same calculations without any
approximation, one gets, for both $q>1$ and $q<1$, $\Delta S^R(\delta,w)/S_{max}\propto
(\delta/2)^q$ for arbitrarily small $\delta$. The observability condition\cite{Lesche} is
ensured. This result is in addition consistent with the fact that $S^R$ is a monotonic
function of $S^T$ which is observable according to the same analysis\cite{Abe02x}. We indeed
have $dS^R=\frac{dS^T}{1+(1-q)S^T} =\frac{dS^T}{\sum_ip_i^q}$. So if $dS^T/S^T\rightarrow 0$,
we also have $dS^R/S^R\rightarrow 0$ for finite $S^R$ and $S^T$. In conclusion, the asymptotic
behaviors of $S^R$ for $w\rightarrow \infty$ and for $\delta\rightarrow 0$ do not commute. In
general, without approximation, $S^R$ should be stable just like $S^T$.

\section{Nonadditive Boltzmann statistics?}
Boltzmann entropy is additive if the product joint probability holds. There is no doubt on
this point. But can this entropy be applied to nonextensive systems with nonadditive energy?
We think that this is possible for microcanonical ensemble, as claimed by
Gross\cite{Gross,Gross1} who and coworkers have treated many systems with long range
interaction or finite size with Boltzmann entropy. This viewpoint is theoretically supported
by R\'enyi nonextensive statistical mechanics\cite{Wang03a} constructed for systems having
nonadditive energy. Since R\'enyi entropy is identical to Boltzmann one for microcanonical
ensemble, R\'enyi statistics is reduced to Boltzmann one and may continue to apply for
nonextensive systems. As a consequence, the statement that Boltzmann statistics is an
extensive theory is not exact because it may work with nonadditive energy.

A point should be clear that Boltzmann entropy does not make any assumption about the
additivity of entropy so that it may be nonadditive. This same statement applies also to some
other additive entropies if we forget the product join probability as a constraint. So a work
based on Boltzmann entropy and using product joint probability to pass from phase space to
non-correlated single body $\mu$-space is not a proof for the applicability of Boltzmann
entropy to nonextensive systems having nonextensive entropy like black hole\cite{Hayw98}. On
the other hand, one should be careful in the case of nonadditive energy $E$ when using both
the conventional definition of thermodynamic temperature $1/T=\frac{\partial S}{\partial E}$
and the product probability, because here $T$ is not intensive any more due to additive $S$.

\section{Conclusion}
We have shown that additive energy should not be used for discussing fundamental topics of
Tsallis or R\'enyi statistics for nonextensive systems at meta-equilibrium. With additive
energy, Tsallis entropy may become additive and R\'enyi entropy nonadditive. So the
equivalence of Tsallis and R\'enyi entropies established on this basis is not true physically.
Equilibrium $N$-body systems with nonadditive energy or entropy should be described by
generalized statistics whose nature is prescribed by the existence of thermodynamic
stationarity. The nonextensive statistics may be Tsallis or R\'enyi one which is naturally
associated with nonadditive energy and with two different temperature definition. The product
joint probability is in this way a natural consequence of the formalism without supposing
additive energy and independence of subsystems. Another interesting point is that R\'enyi
statistics for nonextensive systems becomes Boltzmann one for microcanonical systems. So, from
theoretical viewpoint, Boltzmann entropy is not necessarily associated with extensive systems
and additive energy.

\section*{Acknowledgement}
We would like to thank Profs. D.H.E Gross, J.P. Badiali, B. Jancovici, R. Toral, S. Abe and L.
Velazquez for fruitful discussions and for sending us important references.


\begin{thebibliography}{99}


\bibitem {Conform}
J.L. Cardy, Conformal invariance and statistical mechanics, in {\it Fields, strings and
critical phenomena, Les Houches 1988}, Ed. by E. Br\'ezin and J. Zinn-Justin, North Horland
(1990)p168

\bibitem {Cardy}
J.L. Cardy, and I. Peschel, {\em Nuclear Physics B,\/} {\bf 300}(1988)377

\bibitem {Janco}
B. Jancovici, G. Manificat and C. Pisani, Coulomb systems seen as critical systems :
finite-size effects in two dimensions, {\em J. Stat. Phys.,\/} {\bf 70}(1994)3147

\bibitem {Ruff95}
M. Antoni and S. Ruffo, {\em Phys. Rev. E,\/} {\bf 52}(1995)2361

\bibitem {Ruff01}
J. Barr\'e, D. Mukamel and S. Ruffo, {\em Phys. Rev. Lett.,\/} {\bf 87}(2001)030601

\bibitem {Ruff02}
J. Barr\'e, D. Mukamel and S. Ruffo, {\em Lecture Notes in Physics,\/} {\bf 602}(2002)45-67

T. Dauxois, S. Ruffo, E. Arimondo, and M. Wilkens, {\em Lecture Notes in Physics,\/} {\bf
602}(2002)1-19

\bibitem {Gross}
D.H.E Gross, Geometricla foundation of thermo-statistics, phase transition, second law of
thrmodynamics but without thermodynamic limit, {\em PCCP,} {\bf4}(2002)863-872

\bibitem {Touchette}
H. Touchette, When is a quantity additive and when is it extensive, {\em Physica A,}
{\bf305}(2002)84-88

\bibitem {Gross1}
D.H.E Gross, Micro-canonical statistical mechanics of some nonextensive systems, {\em Chaos,
Solitons $\&$ Fractals,} {\bf 13}(2002)417-430

D.H.E Gross, Nonextensive hamiltonian systems follow Boltzmann's principle not Tsallis
statistics-phase transition, second law of thermodynamics, {\em Physica A,}
{\bf305}(2002)99-105

\bibitem {Tsal88}
C. Tsallis, {\em J. Stat. Phys.,\/} {\bf 52}(1988)479

\bibitem {Abe}
S. Abe, {\em Physica A,\/} {\bf 269}(1999)403-409.

S. Abe, {\em Physica A,\/} {\bf 300}(2001)417;

S. Abe, S. Martinez, F. Pennini and A. Plastino, {\em Phys. Lett.A,\/} {\bf 281}(2001)126;

S. Abe, {\em Phys. Lett. A,\/} {\bf 278}(2001)249

S. Martinez, F. Nicolas, F. Pennini, and A. Plastino, {\em Physica A,\/} {\bf 286}(2000)489,
physics/0003098;

S. Martinez, F. Pennini, and A. Plastino, {\em Phys. Lett. A,\/} {\bf 278}(2000)47;

S. Martinez, F. Pennini, and A. Plastino, {\em Physica A,\/} {\bf 295}(2001)416;

S. Martinez, F. Pennini, and A. Plastino, {\em Physica A,\/} {\bf 295}(2001)246

\bibitem {Toral}
Raul Toral, {\em Physica A,} {\bf 317}(2003)209-212, cond-mat/0106060

\bibitem {Velazquez}
L. Velazquez and F. Guzman, {\em Phys. Rev. E,\/} {\bf 65}(2002)046134

\bibitem {Raggio}
G.R. Guerberoff and G.A. Raggio, {\em J. Math. Phys. ,\/}{\bf 37},1776(1996);

G.R. Guerberoff, P. A. Pury and G.A. Raggio, {\em J. Math. Phys. ,\/}{\bf 37},1790(1996);

G.A. Raggio, {\em Equivalence of two thermostatistical formalisms based on the
Havrda-Charvat-Dar\'oczy-Tsallis entropies,\/} eprint : cond-mat/9908207;

G.A. Raggio, {\em On equivalence of thermostatistical formalisms,\/} eprint : cond-mat/9909161

\bibitem {Beck01}
C. Beck, {\em Phys. Rev. Lett.,\/} {\bf 87}(2001)180601

\bibitem {Wang02}
Q.A. WANG, {\em Euro. Phys. J. B,} {\bf26}(2002)357

\bibitem {Wang02a}
Q.A. WANG, L. Nivanen, A. Le M\'ehaut\'e and M. Pezeril, {\em J. Phys. A,} {\bf 35}(2002)7003

\bibitem {Wang02b}
Q.A. WANG, {\em Phys. Lett. A,} {\bf 300}(2002)169

\bibitem {Wang02c}
Q.A. WANG and A. Le M\'ehaut\'e, {\em J. Math. Phys,} {\bf 43}(2002)5079

\bibitem {Wang02d}
Q.A. WANG and A. Le M\'ehaut\'e, {\em Chaos, Solitons $\&$ Fractals,} {\bf 15}(2003)537-541

\bibitem {Letters}
S. Abe, A.K. Rajagopal, A. Plastinos, V. Lotora, A. Rapisarda and A. Robledo, Letters to the
Editor : Revisiting disorder and Tsallis statistics, {\em Science,\/} {\bf 300}(2003)249-251


\bibitem {Cho}
A. Cho, A fresh take on disorder, or disorderly science? {\em Science,\/} {\bf 297}(2002)1268


\bibitem {Nauenberg}
M. Nauenberg, A critique of q-entropy for thermal statistics, {\em Phys. Rev. E,\/} {\bf
67}(2002)036114


\bibitem {Penni}
F. Pennini, A.R. Plastino and A. Plastino, {\em Physica A,\/} {\bf 258}(1998)446

C. Tsallis, R.S. Mendes and A.R. Plastino, {\em Physica A,\/} {\bf 261}(1999)534;

Silvio R.A. Salinas and C. Tsallis, {\em Brazilian Journal of Physics(special issue:
Nonadditive Statistical Mechanics and Thermodynamics),\/} {\bf 29}(1999).

\bibitem {Almeida}
M.P. Almeida, {\em Physica A,\/} {\bf 300}(2002)424-432

M.P. Almeida, An additive variant of Tsallis generalized entropy, cond-mat/0208064

A.B. Adib, A.A. Moreira, J.S. Andrade and M.P. Almeida, Tsallis thermostatistics for finite
systems : a Hamiltonian approach, cond-mat/0204034

\bibitem {Abe3}
S. Abe, {\em Physica A,\/} {\bf 305}(2002)62-68

\bibitem {Reny66}
A. R\'enyi, {\em Calcul de probabilit\'e,\/}(Paris, Dunod, 1966)P522.

\bibitem {Tsal02}
C. Tsallis and A.M.C. Souza, {\it Constructing a statistical mechanics for Beck-Cohen
superstatistics,} cond-mat/0206044

\bibitem {Abe2}
S. Abe, {\em Phys. Lett. A,\/} {\bf 278}(2001)249-254

\bibitem {Vives}
E. Vives and A. Planes, {\em Phys. Rev. Lett.,\/} {\bf 88}(2002)12061

\bibitem {Lenzi}
E.L. Lenzi, R.S. Mendes and L.R. da Silva, {\em Physica A,\/} {\bf 280}(2000)337

\bibitem {Bashkirov}
A.G. Bashkirov and A.V. Vityazev, {\em Physica A,\/} {\bf 277}(2000)136


\bibitem {Wang03a}
Q.A. WANG, { \it Power law distribution of R\'enyi entropy for equilibrium systems having
nonadditive energy,} cond-mat/0304146

\bibitem {Abe5}
S. Abe, {\em Phys. Rev. E,\/} {\bf 63}(2001)061105

\bibitem {Wang01}
Q.A. Wang, {\em Chaos, Solitons $\&$ Fractals,}{\bf 12}(2001)1431; Erranta :
cond-mat/0009343

\bibitem {Lesche}
B. Lesche, {\em J. Stat. Phys.,} {\bf27}(1982)419

\bibitem {Abe02x}
S. Abe, {\em Phys. Rev. E,} {\bf 66}(2002)046134

\bibitem {Hayw98}
Sean A. Hayward, {\em Class. Quant. Grav.,\/} {\bf 15}(1998)3147


\end{thebibliography}
\end{document}